%% file: main.tex
\documentclass[twocolumn, tighten]{aastex63}

\usepackage{float}

\usepackage[figuresright]{rotating}
\usepackage{makecell}

\newcommand{\mytilde}{\raise.17ex\hbox{$\scriptstyle\sim$}}

\graphicspath{{./}{figures/}}

\shorttitle{Embedded Protostars in Ophiuchus}
\shortauthors{Encalada et al.}

\begin{document}

\title{870 MICRON DUST CONTINUUM OF THE YOUNGEST PROTOSTARS IN OPHIUCHUS}

\correspondingauthor{Frankie J. Encalada}
\email{fje2@illinois.edu}

\author{Frankie J. Encalada}
\affil{Department of Astronomy, University of Illinois, 1002 W. Green St., Urbana, IL 61801, USA \\}

\author{Leslie W. Looney}
\affil{Department of Astronomy, University of Illinois, 1002 W. Green St., Urbana, IL 61801, USA \\}

\author{John J. Tobin}
\affil{National Radio Astronomy Observatory, 520 Edgemont Rd., Charlottesville,VA 22903, USA}

\author{Sarah I. Sadavoy}
\affil{Department of Physics, Engineering and Astronomy, Queen’s University, 64 Bader Lane, Kingston, ON, K7L 3N6, Canada}

\author{Dominique Segura-Cox}
\affil{Center for Astrochemical Studies, Max Planck Institute for Extraterrestrial Physics, 85748 Garching, Germany \\}

\author{Erin Cox}
\affil{Center for Interdisciplinary Exploration and Research in Astrophysics (CIERA), 1800 Sherman Avenue, Evanston, IL 60201, USA}

\author{Zhi-Yun Li}
\affil{Department of Astronomy, University of Virginia, Charlottesville, VA 22904, USA}

\author{Giles Novak}
\affil{Center for Interdisciplinary Exploration and Research in Astrophysics (CIERA), 1800 Sherman Avenue, Evanston, IL 60201, USA}

\begin{abstract}

We present a 0.15$^{\prime\prime}$ resolution (21 au) ALMA 870 $\mu$m continuum survey of 25 pointings containing 31 young stellar objects in the Ophiuchus molecular clouds. Using the dust continuum as a proxy for dust mass and circumstellar disk radius in our sample, we report a mean mass of 2.8$^{+2.1}_{-1.3}$ and 2.5$^{+9.2}_{-1.1}$ M$_{\oplus}$ and a mean radii of 23.5$^{+1.8}_{-1.2}$ and 16.5$^{+2.8}_{-0.9}$ au, for Class I and Flat spectrum protostars, respectively.  In addition, we calculate the multiplicity statistics of the dust surrounding young stellar objects in Ophiuchus. The multiplicity fraction (MF) and companion star fraction (CSF) of the combined Class I and Flats based solely on this work is 0.25 $\pm$ 0.09 and 0.33 $\pm$ 0.10, respectively, which are consistent with the values for Perseus and Orion. While we see clear differences in mass and radius between the Ophiuchus and Perseus/Orion protostellar surveys, we do not see any significant differences in the multiplicities of the various regions. We posit there are some differences in the conditions for star formation in Ophiuchus that strongly affects disk size (and consequently disk mass), but does not affect system multiplicity, which could imply important variation in planet formation processes.

\end{abstract}


\section{Introduction} \label{sec:intro}

Molecular clouds form dense cores and eventually protostars when gravity overcomes the internal support mechanisms, e.g., magnetic fields, thermal pressure, and turbulence \citep[e.g.,][]{mckee2007}. Each core can form one star or a small system of stars. Observationally, embedded stars are characterized by their spectral energy distributions (SEDS), specifically between $\sim$2 $\mu$m to 24 $\mu$m \citep{lada1987}, as well as by their bolometric temperature. 

The youngest stage of this evolution is characterized by the Class 0 protostars, which are deeply embedded, have strong bipolar outflows, and are accreting from the surrounding medium \citep{andre1993}. After $\mytilde10^{5}$ yr \citep{mike2014}, the star reaches the Class I stage, which features continued accretion onto the central protostar and a significant reduction in the envelope that surrounds the system. Due to variations in inclination, the observational differential from Class 0 to I is subtle with many authors using a $T_{bol}$ $<$ 70 K for Class 0 sources \citep[e.g.,][]{mike2014}. As the protostars evolve, they become less embedded and detectable at micron wavelengths, as their SEDs become more flattened and eventually their slopes turn negative by the Class II stage (T$_{bol}$ $<$ 650 K). After $\mytilde10^7$ yr the disks have completely dissipated.  In addition, there are often transitional sources between Class I and II called Flat Spectrum sources, named for their flat SEDs (in $\lambda$F$_\lambda$).

Important to the evolution of the forming star and the formation of planets is the circumstellar disk around the the young stars themselves.  Several open questions remain about these disks, including their mass and size at the protostar stage and how their initial properties vary in different star forming regions. In the last few years, with the increased sensitivity of interferometers, we have made the largest leaps in characterization of the youngest disks, which previously had only been constrained in small numbers \citep[e.g.,][]{looney2000}.

Recently, the VLA/ALMA Nascent Disk and Multiplicity (VANDAM) survey in Perseus identified a total of 18 Class 0 disk candidates with radii less than 30 au \citep{dom2018}.  However, the largest survey of young disks to date was just completed by the VANDAM Orion survey \citep{tobin2020}, which observed 328 protostars ranging from Class 0 to Flat Spectrum in the Orion molecular cloud.  In that sample, they found that the mean dust disk mass decreased with evolutionary state, which is arguably expected due to accretion onto the protostar and possible grain growth.  They also found that the mean disk dust radii decreased with evolutionary state, which is harder to understand for simple disk growth from rotating cores.  In addition, the Orion VANDAM survey presented dust disk masses distributions in comparison to other star formation regions.  They concluded that the masses in Orion are comparable to those in Perseus and Taurus, but are more massive than the disk masses in Ophiuchus.

Multiplicity is very common in main sequence stars with $\sim$50\% of nearby solar-mass stars being part of multiple systems \citep[e.g.,][]{Raghavan2010}, with the multiplicity fraction increasing with stellar mass \citep[e.g.,][]{Duchene2013}. However, multiplicity also evolves with stellar age as younger objects have higher multiplicity \citep[e.g.,][]{Connelley2008}.  Indeed, multiplicity is found to commonly occur at the earliest, deeply embedded stage of star formation \citep[e.g.,][]{looney2000, tobin2016}, which is where multiplicity formation originates \citep[e.g.,][]{tobin2018}.  Probing the multiplicity of these systems requires detection of the dust structures surrounding the protostar rather than the detection of the star itself, as is done in more evolved stellar systems.

The largest survey of multiplicity in the youngest protostars to date are the VANDAM surveys of Perseus \citep{tobin2016, tobin2018} and Orion (Tobin et al., in prep.).  In both cases, they find a $\sim$40\% overall multiplicity fraction with a companion star fraction of $\sim$70\%.  However, the most interesting result was a bimodal nature of the binary separations: a peak at $\sim$75 au and another one at $\sim$3000 au.  These peaks are consistent with binary formation from disk and turbulent fragmentation, respectively.   

Ophiuchus has over 300 young stellar objects (YSOs) identified by the Spitzer From Molecular Cores to Planet-Forming Disks project (C2D) \citep{evans2009} and others \citep[e.g.,][]{enoch2009, zurlo2020}.  \cite{cox2017} surveyed 63 disks (mostly Class II) in Ophiuchus using ALMA Band 7 dust continuum observations, which showed clear variations in disk mass and size between single and multiple systems. In addition, \cite{Cieza2019} surveyed a larger sample (using less restrictive selection criteria than our sample with respect to previous detections) of Ophiuchus sources (147 sources with 133 detected, including Class I and Flat Spectrum protostars) in the ALMA Band 7 dust continuum, which suggested that only 1/3 of the sample had disk masses large enough to form planetary systems with gas giants, i.e. $>$ 10M$_\Earth$.  Finally, \cite{sadavoy2019} surveyed 37 Ophiuchus systems (including Class I and Flat Spectrum protostars) in the ALMA Band 6 dust polarization continuum, and also used the Stokes I continuum to study multiplicity.  They found MF = 0.29 and CSF = 0.41 for the combined Class 0 and Class I population, similar to those found in Perseus \citep{tobin2016}. Finally, \cite{zurlo2020} surveyed the \cite{Cieza2019} sample with deep IR imaging to reveal 20 new binary systems and 2 new triple system, and also showing that the binary systems have lower disk masses than those in single systems. As in the case of the above four studies, we adopt the common distance of 140 pc for Ophiuchus \citep{ortizleon2018}. 

In this paper, we present dust mass observations of 31 of the youngest protostars in Ophiuchus using Band 7 (870 $\mu$m wavelength) of ALMA. In Section 2, we discuss our sample selection. In Section 3, we explain the observations. In Section 4, we discuss the results, especially in comparison to the Orion star formation region. And in Section 5, we summarize.

\section{Sample Selection} \label{sec:sample}

The main objective for this sample was to complement the sources from our previous survey of protostellar systems in Ophiuchus that concentrated on Class II/III systems of \cite{cox2017}. The reader is directed towards their paper for specifics, but briefly we required that sources must be detected with a signal-to-noise ratio $>$3 in both the four IRAC bands and the 24 $\mu$m MIPS bands, and also the sources must be detected with a signal-to-noise ratio $>$2 in the 70 $\mu$m MIPS band. 
This selection criteria was used to minimize the impact from misclassified YSO sources and extragalactic sources, as well as to maximize submillimeter detections.
Ultimately, the most deeply embedded sources \citep[e.g., 22 Class 0 and I sources from][]{enoch2009} were removed from the \cite{cox2017} sample to focus more on the disk-only sources. Our new observations targeted the 22 excluded sources plus the remaining 6 Enoch sources that did not meet the previous requirements (possible younger sources). These 6 sources are Oph-emb 5, Oph-emb 7, GY30, CRBR 85, GY269, and GY91. We did not observe sources that had been observed at our resolution or higher (e.g., IRAS 16293-2422, VLA 1623-2417, and GY214) but we did include them in the multiplicity analysis. This left 25 sources in total for our observations. Table \ref{tb:sources} lists the Class 0/I sources observed, which with binary systems and a potential new source brings the total number of objects to 31. Note that with the exclusion of the sources observed in other programs, there is only one Class 0 source remaining in our sample.

Of our 31 sources, 22 overlap (including binaries) with the lower resolution ($\sim$0.25$^{\prime\prime}$) Band 6 polarization observations of \cite{sadavoy2019}. The sources Oph-emb 23, 24, and 27 were excluded from \cite{sadavoy2019}. In addition, the $\sim$0.2$^{\prime\prime}$ Band 6 continuum observations of \cite{Cieza2019} contain 20 sources that overlap with ours and 10 that do not (sources they did not observe: Oph-emb 1, 7, 8, 10A, 10B, 13, 14A, 14B, 21, and 26B).

\section{Observations and Data Reduction} \label{sec:data}

We observed our sources with 25 pointings of the continuum-only dual configuration from our ALMA Band 7 project 2015.1.00741.S. The high resolution observations in the C40-6 configuration occurred between 23:44:01 UTC on 2016 August 29th to 00:28:22 UTC on 2016 August 30th. The baselines ranged from 15 to 612 meters, giving the largest angular scale recoverable as 1.4$^{\prime\prime}$. During the observations, the precipitable water vapor was 0.80 $\pm$ 0.05 mm. The average time on source was 60 seconds. The low resolution observations in the C36-2/3 configuration occurred between 04:53:36 to 05:31:35 UTC on 2016 April 22nd. The baselines ranged from 15 to 460 meters, providing the largest angular scale recoverable as 7.2$^{\prime\prime}$. During the observations, the precipitable water vapor was 0.64 $\pm$ 0.02 mm. The average time on source was 30 seconds. In both configurations, we used the quasar J1517-2422 as a bandpass calibrator and J1625-2527 as phase calibrator. The flux calibrator was J1633-2557. We assume the absolute flux calibration accuracy to be $\sim$10\%, in accordance with the ALMA handbook \citep{almahandbook2018}. Therefore, only statistical uncertainties are used for the rest of the paper.

The data were reduced using the CASA, Common Astronomy Software Applications \citep{casa}, version 4.7.0 pipeline, and the self-calibration and mapping were performed using CASA 5.1.1. The continuum maps were made using Briggs weighting with a robust parameter of 0.5 (and 0.8 in two cases to recover larger-scale structure), yielding a typical synthesized beam of 0.15$^{\prime\prime} \times$0.11$^{\prime\prime}$ full-width at half-maximum (FWHM). However, this does not hold true for Oph-emb 04 and 22. Both of those had over 75\% of their observations flagged. This led to larger than expected beams.

We followed standard self-calibration techniques for all sources. To briefly summarize, at most we performed three rounds of phase-only self-calibration with gain calibration solutions intervals ranging from the entire scale length, to 30 seconds, and finally 15 seconds (with single interval integrations of 2.02 seconds). We chose the gain calibration solution with the most consistent solutions that decreased the noise or at least increased the overall S/N. We conclude that the point-sources saw marginal improvements between self-calibration rounds while the extended sources saw the most benefits.

\section{Results} \label{sec:results}

\begin{figure*}[h]
    \centering
    \includegraphics[width=1.0\textwidth]{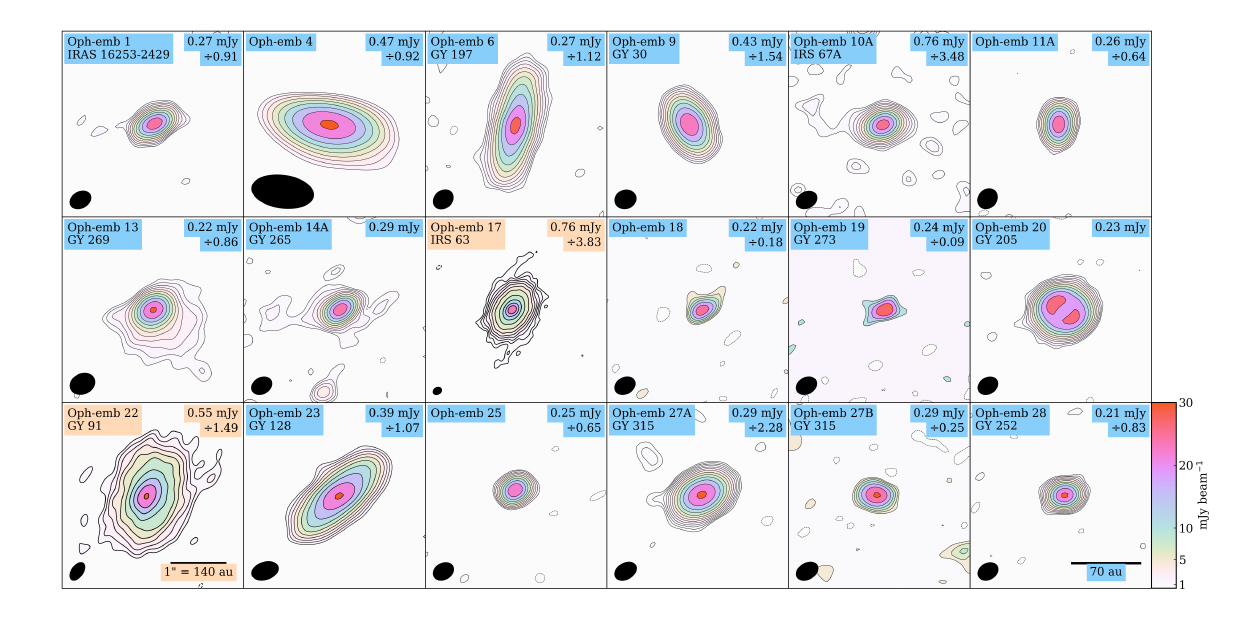}
    \caption{Maps of all the well-resolved sources in our sample. The root-mean-square (RMS) of the noise is in the top right corner. The flux in each image has been scaled by the amount next to the division symbol in the top right corner. Images that share the same text background color also share the same scalebar; IRS 63 and GY 91 are zoomed out twice as far as the other sources (although GY 91 suffers from the same issue as Oph-emb 04). The beam is in the bottom left corner and the values can be found in Table \ref{tb:sources}. Dotted contours are at -$3\sigma$ and solid contours are at $3\sigma \times \sqrt[2]{2^{n}}$ for $n \in \mathbb{W}$ (i.e., 3$\sigma$ [-1, 1, 1.41, 2, 2.82, 4, ...])}
    \label{fig:resolved}
\end{figure*}

\begin{figure*}[h]
    \centering
    \includegraphics[width=1.0\textwidth]{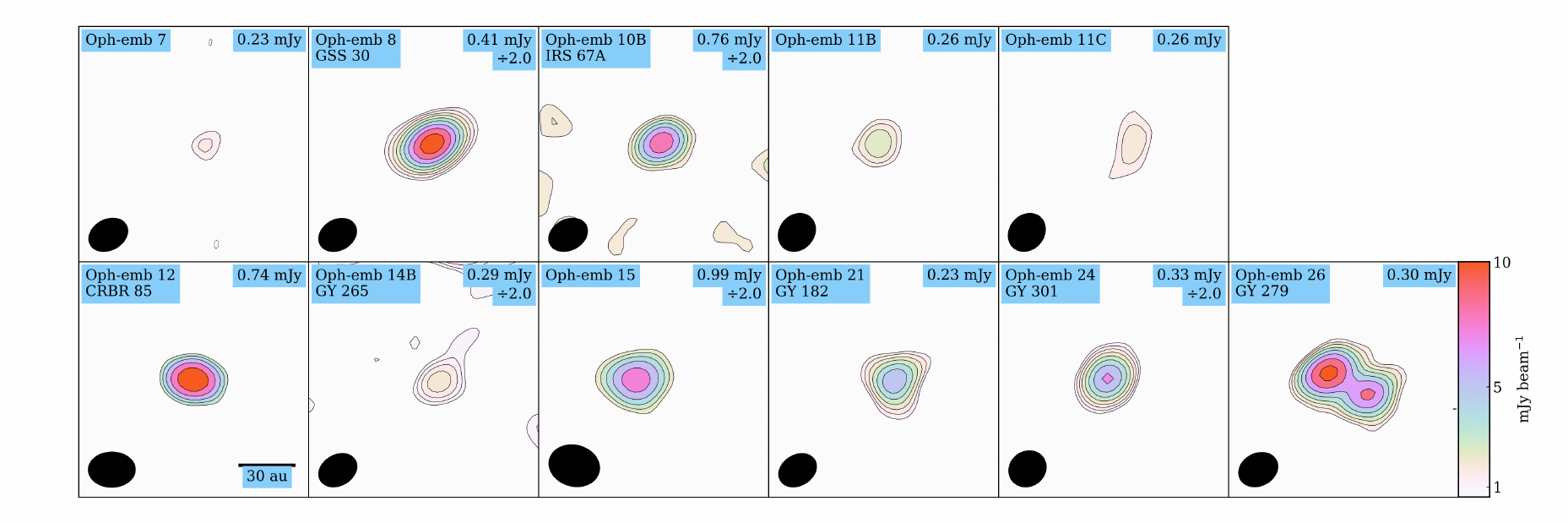}
    \caption{The less resolved sources from our sample. The smaller peak in the bottom right corner of Oph-emb 26 is assumed to be a new source. Otherwise, same as Figure \ref{fig:resolved}.}
    \label{fig:unresolved}
\end{figure*}

\begin{figure*}[h]
    \centering
    \includegraphics[width=0.65\textwidth]{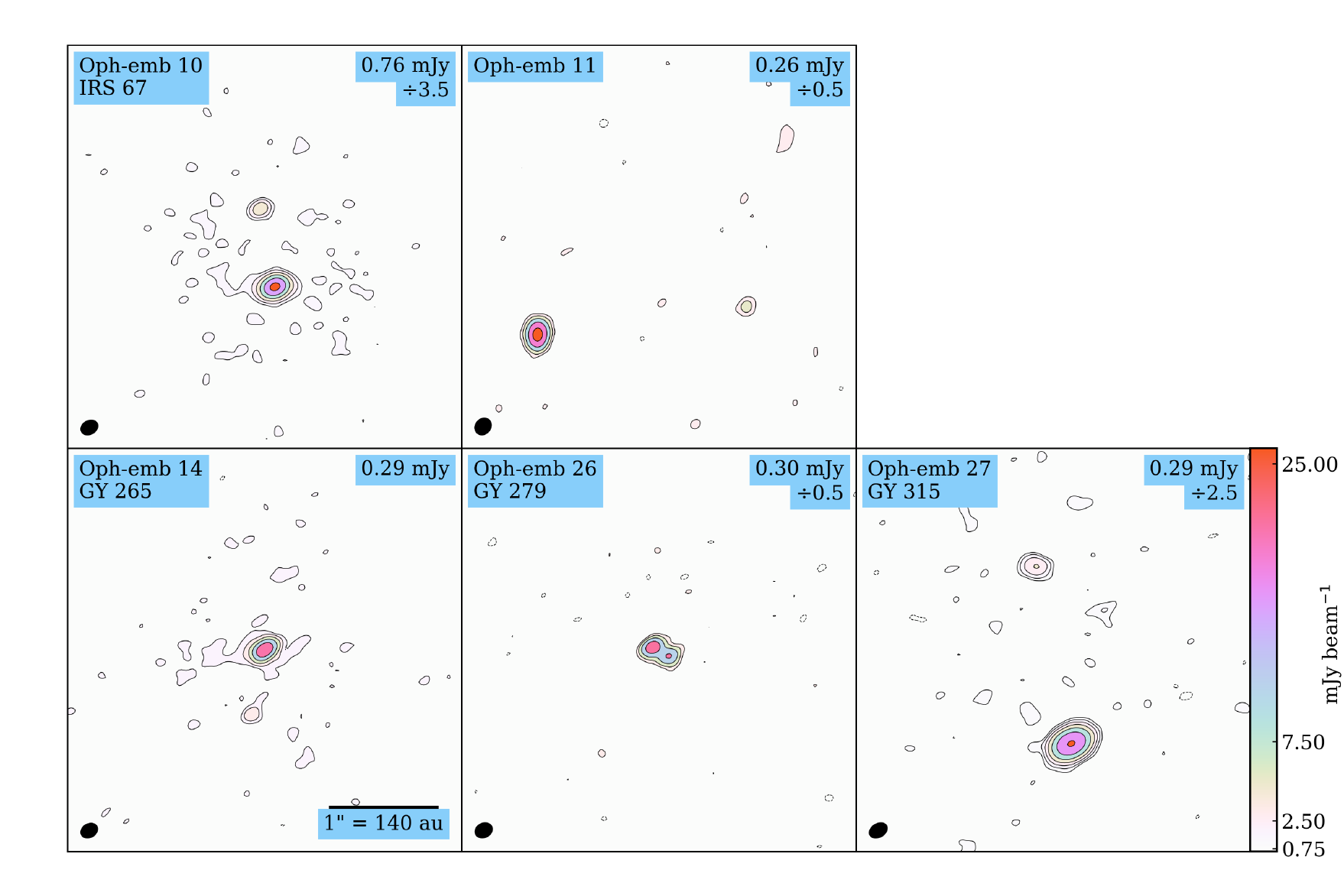}
    \caption{The multiplicity sources from our sample. Dotted contours are at -$3\sigma$ and solid contours are at $3\sigma \times \sqrt[2]{2^{2n}}$ for $n \in \mathbb{W}$. Otherwise the same as Figure \ref{fig:resolved}.}
    \label{fig:multiples}
\end{figure*}

We detected 24 of the 25 sources at the pointing centers observed in the Band 7 dust continuum. The non-detection of Oph-emb 5 is consistent with the findings of \cite{sadavoy2019}. We also detected 6 additional companions in our pointings, bringing the total to 31 sources detected. The pointing maps are shown in Figures \ref{fig:resolved} and \ref{fig:unresolved}, split into the most resolved and least resolved sources respectively.  Note that the disk sizes and morphologies are quite varied. In Figure \ref{fig:multiples}, we highlight the multiple systems.

\begin{figure*}[h]
    \centering
    \includegraphics[width=0.4\textwidth]{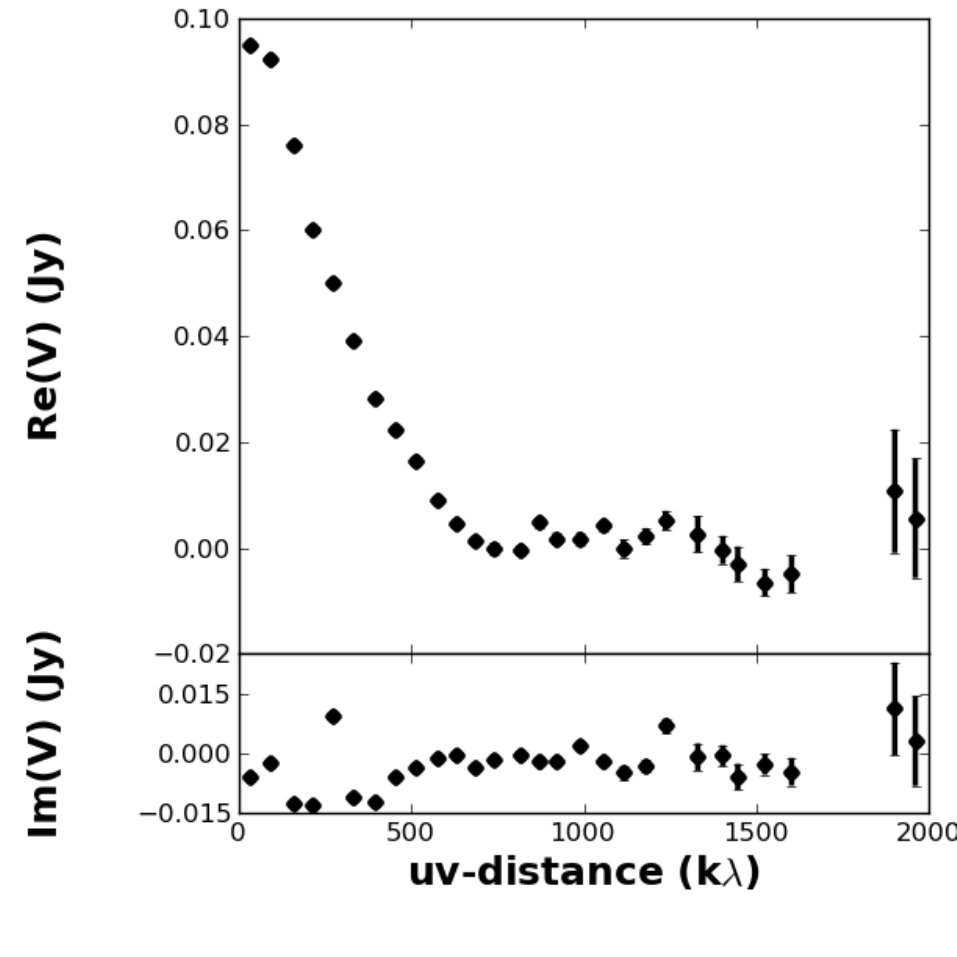}
    \caption{An example of the real and imaginary components of the binned visibilities of a resolved source (Oph-emb 9) in our sample.  This is shown solely to demonstrate the typical shape of the curve that shows evidence of a disk. The imaginary component (bottom) is expected to be around zero.  The real component (top) shows the extended envelope on the left (the steady decline) with the resolved disk near the middle (the multiple bounces).}
    \label{fig:uvplot_09}
\end{figure*}

One of the main objectives of this study is to determine circumstellar disk properties.  In the majority of cases, the dust emission is resolved.  As we are probing size scales expected for disks in Class I systems \citep[e.g.,][]{tobin2012,dom2016,tobin2020}, we made {\it u,v} plots of the Real and Imaginary components of the visibilities, see Figure \ref{fig:uvplot_09} for an example.  In all cases, the visibilities were consistent with disks \cite[e.g.,][]{dom2018}, often with residual envelope emission.  From this point forward, we will refer to these objects as disks instead of disk candidates.  Of course, without Keplerian-like velocity fields (which were not observed in these observations), we can not confirm their disk nature.

For every detected source in our sample, we used the CASA task \texttt{imfit} to fit a Gaussian within an elliptical region centered on the object. 
The task itself simplifies the Gaussian fitting process, and returns useful quantities such as the properties of the ellipse (major, minor, and position angle) and the flux density, which we use to measure disk properties.
Our observations resolve out the majority of the envelope emission, so a single Gaussian provides the best fit to the images.

We follow the procedure of \cite{tobin2020} for estimating disk properties. Although there are many methods available for estimating disk radii, Gaussian image fits provide the simplest and most easily compared. We adopt the disk radius as 2$\sigma$ of the deconvolved major axis from the Gaussian fit.
For consistency of fit, the size of the ellipse was varied between the 3$\sigma$ and 6$\sigma$ contour. This criteria ensures invariance to initial fit parameters and were applied so long as reasonable (e.g., no other object in the way).  We report 3$\sigma$ contour-sized fit results.
Oph-emb 26 was fit using two Gaussians simultaneously due to the proximity of the binary. One source (Oph-emb 05) was a non-detection so only its 3$\sigma$ upper limit peak flux is listed. The residuals were all checked for any visually bad fits. Some source fits were point sources, so their major and minor axes are not given in Table \ref{tb:sources}.  

We estimate the dust mass of the compact circumstellar material using a series of broad, simple assumptions and the Gaussian fits. Following \citep{tobin2020}, assuming isothermal and optically thin emission (both of which are very idealized for these sources at this wavelength), we can estimate the mass from,

\begin{equation}
M_{dust} = \frac{d^{2} F_{\nu}}{B(T_{dust}) \kappa_{\nu}}
\label{eq:mdust}
\end{equation}

\noindent where $d$ is the assumed distance (140 pc) to Ophiuchus, $F_{\nu}$ is the total observed flux, $B_{\nu}$ is the Planck function, and $T_{dust}$ is the dust temperature. The grain opacity, $\kappa_{\nu}$, is taken as 1.84 cm$^{2}$ g$^{-1}$ at 870 $\mu$m \citep[e.g.,][]{OH1994}. This is done to be consistent with the VANDAM Orion protostellar survey, but we acknowledge that this dust opacity likely overestimates the dust mass because we expect higher densities and more extensive grain growth in disks compared to the OH94 models, which are applicable to the envelopes of protostars. Likewise, to be consistent with the VANDAM Orion protostellar survey \citep{tobin2020}, we also adopt a dust temperature based on modeling that is dependent on the bolometric luminosity,

\begin{equation}
T_{dust} = T_{0} \left(\frac{L_{bol}}{1\ L_{\odot}}\right)^{\frac{1}{4}}
\end{equation}

\noindent where $T_0$ = 43 K and $L_{bol}$ is from \cite{enoch2009}, given in Table \ref{tb:sources}.   In the case of multiple components, we used the same $L_{bol}$ for the components for simplicity.  The calculated dust masses are also listed in Table \ref{tb:sources}, ranging from 0.4 to 89.3 M$_\Earth$, with 0.1 M$_\Earth$ at 4$\sigma$. 
The uncertainty in the calculated dust mass is a factor of a few due to the assumptions of optically thin material, dust opacity, and constant temperature.

\section{Discussion} \label{sec:discuss}

One of the main motivators in these observations is to compliment a survey by \cite{cox2017} of the Ophiuchus region. A full survey of the Orion region Class 0 and Class I protostars, called the VANDAM survey was completed \citep{tobin2020}, which provides a complete look at the young protostars in the Orion region. The use of resolved imaging of the disk can be combined with $L_{bol}$ and $T_{bol}$ to make links between disk masses and radii and evolutionary stage. The Orion VANDAM survey reveals a slight decrease in disk mass with evolution, which is expected from disk evolution and accretion, but on the other hand, they also showed a decrease in disk radii with evolution. The latter may be due to some intrinsic initial conditions of the disk formation, some evolution in dust grains, variation in the environment, or differences in the underlying stellar populations.  We add our observations to the Ophiuchus observations used in the \cite{tobin2020} comparisons, which includes disk properties and multiplicity (which will be an upcoming Orion VANDAM paper).

\subsection{Disk Properties}

\begin{figure*}[h]
    \centering
    \includegraphics[width=0.7\textwidth]{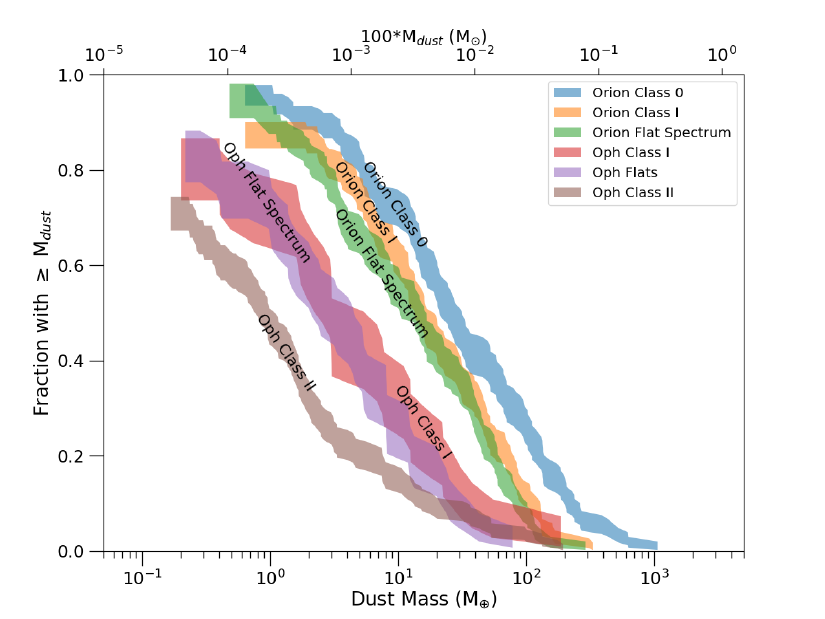}
    \caption{Cumulative distribution from \cite{tobin2020} with data added from this paper (and the non-overlapped sources from \cite{williams2019}). The key difference is that the Ophiuchus Class I and flat spectrum masses are even lower than the original figure.}
    \label{fig:ysocumdist}
\end{figure*}

In Figure \ref{fig:ysocumdist}, we plot the cumulative distribution function of disk masses in Orion and Ophiuchus, as in \cite{tobin2020} (Figure 14, top left, but without the Perseus sample), where the Orion source masses are from that paper and the Ophiuchus source masses are from this paper and \cite{williams2019}. The \cite{williams2019} masses were calculated using a constant, and typically, lower temperature of 20 K and a higher dust opacity of  
2.25 cm$^2$ g$^{-1}$ (extrapolated to  225 GHz). We compare our derived masses with the 18 sources that overlapped with \cite{williams2019}. In most cases, our masses (corrected to use the same distance of 139.4 pc) are lower than the \cite{williams2019} mass estimates by an average of 50\%.  This is mainly due to our higher temperature estimates; the average temperature based from Equation 2 was 40 K.

With the addition of our masses, the overall shape of the curve is similar to \cite{tobin2020}, but the separation between the disk masses of the Orion and Ophiuchus protostars is even more pronounced. Indeed, the difference in masses reinforces one of the outstanding questions about this relation, and discussed in detail by \cite{tobin2020}: \textit{why is there a discrepancy between the disk masses in Ophiuchus compared to the disk masses in Orion?}  As discussed in \cite{tobin2020}, disk masses in Taurus, Perseus, and Orion are comparable, but Ophiuchus, being on the lower end, remains an outlier. Indeed, the masses in our sample are derived using the same procedure and the same ALMA band (Band 7) as \cite{tobin2020}, and yet the difference between the Class 0/I/F protostars has only increased.  Thus, the discrepancy does not appear to be based on the assumptions across the bands; Ophiuchus disks are much less massive than Orion disks.  

As suggested by \cite{tobin2020} and \cite{sadavoy2019}, there are possible issues with some sources being misidentified as younger classes due to foreground extinction, etc., but the Oph disk sample in Figure \ref{fig:ysocumdist} is obviously lower in mass than the other regions. Some possibilities for this discrepancy are: 1) due to the lower number of protostellar objects in Ophiuchus, since we are sampling a majority of older Class I sources (which is somewhat consistent with the very low number of Class 0 in Ophiuchus); 2) a fundamental difference in the dust opacity between the two regions, which is not generally seen  \citep[e.g.,][]{PlanckXI}; or 3) differences in the the initial conditions of the disks, which would imply differences in the star formation process between the two clouds.  Of the three, we propose that the differences in the initial conditions of the disks to be the most likely since our masses make the discrepancy more pronounced even between the Class I in Ophiuchus and the Flat Spectrum sources in Orion.

Finally, we used the cumulative distributions and survival analysis with the Kaplan-Meier estimator as implemented in the Python package \texttt{lifelines} to fit a Gaussian of the mass cumulative distribution function (CDF). One of the benefits of survival analysis is the ability to incorporate non-detections. We estimate the disk masses for our sources in Figure \ref{fig:ysocumdist}, i.e., excluding the \cite{williams2019} sources. We find a mean mass of 2.8$^{+2.1}_{-1.2}$ M$_\Earth$ for Class I and 2.5$^{+9.2}_{-1.1}$ M$_\Earth$ for Flats.  As expected, these are significantly less than the masses in Orion of 14.9$^{+3,8}_{-2.2}$ and 11.6$^{+3.5}_{-1.9}$ M$_\Earth$, respectively, from \cite{tobin2020}.

Following \cite{tobin2020}, we also estimate the mean radii of our disks in Table \ref{tb:sources} by again using the cumulative distributions and fitting a Gaussian radius CDF for our Class I and Flat sources. This allows us to better include the unresolved sources in the estimate.  We find mean radii of 23.5$^{+1.8}_{-1.2}$ and 16.5$^{+2.8}_{-0.9}$ au, respectively.  This is in comparison to \cite{tobin2020}, which finds 44.9$^{+5.8}_{-3.4}$, 37.0$^{+4.9}_{-3.0}$, and 28.5$^{+3.7}_{-2.3}$ au for Class 0, I, and Flats, respectively.  We only have one Class 0 source (Oph-emb 1), which has an image Gaussian fit radius of 15 au, so we did not use the CDF method discussed above.  We note that 15 au is much smaller than the mean Class 0 radii in Orion, but with so few Class 0 sources in Ophiuchus it is difficult to draw any firm conclusions.  However, overall, the disk radii and disk mass for Ophiuchus protostars seem to be smaller than those in Orion without any clear trend with evolution, whereas Orion and Perseus are consistent with each other.

\subsection{Multiplicity}

To find multiplicity in our sample, we visually inspected our maps out to the FWHM of the primary beam of ALMA.  We define a companion protostar as any source with a S/N $\gtrsim$ 4 or if the \texttt{imfit} task required two sources to obtain an acceptable residual map (i.e., Oph-emb 26). We infer source multiplicity from millimeter emission.  The detection of a source is dependent on 1) dust structures surrounding the source, 2) dust opacity and temperature, 3) resolution to resolve close systems, and 4) the overall sensitivity of the observations. We detect a total of 6 companions in our observations. One of which is possibly a new multiple: Oph-emb 26. We measured separations using the \texttt{imfit} central positions as listed in Table \ref{tb:sources}. We combine our observations with those of \cite{Cieza2019}, adding all detected Class I/Flat sources. We use a criteria for triple systems wherein we measure distances from the primary, obtaining only two separation distances. Overall, the total sample includes 14 measured separations for young multiple systems in Ophiuchus, see Table \ref{tb:binaries}. Of the 14 total, 6 are from this paper, 4 are from \cite{Cieza2019}, and the last 4 are from the two triple systems, VLA 1623 (\cite{looney2000}, \cite{harris2018}, \cite{hsieh2020}) and IRAS 16293 \citep{wootten1989}.

The best published analysis of multiplicity in protostars to date is from the VANDAM Perseus collaboration \citep{tobin2016}. Although having a shorter wavelength (9~mm) and wider field of view with the VLA component, the VANDAM Perseus survey has comparable spatial resolution (15 au) but lower mass sensitivity \citep[$\sim$10 M$_\Earth$:][]{Tychoniec2020}, which still allows us to compare the inferred multiplicity between two star formation regions that have a discrepancy in disk masses. (The \cite{Cieza2019} multiples are at Band 6 with a larger FWHM.)

The metrics commonly used for multiplicity are the multiplicity fraction (MF) and companion star fraction (CSF), which can be considered the probability that any given system in the sample has companions (MF) and the average number of companions (CSF).  The quantities are defined by

\begin{equation}
MF = \frac{B+T+Q+...}{S+B+T +Q+...}
\end{equation}

\noindent and

\begin{equation}
CSF = \frac{B+2T+3Q+...}{S+B+T +Q+...},
\end{equation}

\noindent where S, B, T, and Q refer to the number of single, binary, triple, and quadruple systems.

Uncertainties for the two metrics are derived assuming binomial distributions as $\sigma_R = (R (1-R)/N_{sys})^{0.5}$, where  R is either the MF or CSF value and $N_{sys}$ is the number of systems.  With those assumptions, the values we obtain for only the combined Class I and F sources in Ophiuchus surveyed in this paper are MF = 0.25 $\pm$  0.09 and CSF = 0.33 $\pm$ 0.10, see Table \ref{tb:mfncsf}.  These are consistent with the metrics of the combined Class I and F sample of Perseus with a separation range of 20 to 1000 au from Tobin et al. (2021, in preparation), which reported MF = 0.28 $\pm$ 0.08 and CSF = 0.28 $\pm$ 0.08, see Table \ref{tb:mfncsf}. In addition, these are all consistent with the larger VANDAM Orion sample, which reports MF = 0.16 $\pm$  0.02 and CSF = 0.17 $\pm$ 0.03 (Tobin et al., 2021 in preparation), which is at the same wavelength band but with lower resolution (45 au) and sensitivity ($\sim$1 M$_\Earth$). Furthermore, applying the same analysis to the larger sample of protostellar Class I and F objects in Ophiuchus, including \cite{Cieza2019} sources, gives an MF of 0.11 $\pm$ 0.04 and an CSF of 0.14 $\pm$ 0.04 (see Table \ref{tb:mfncsf}), which is even more consistent with the Orion sample. This may be due to the less restrictive sample selection criteria of Tobin et al. (2021, in preparation) and \cite{Cieza2019}, where more C2D sources were included in the former and all C2D sources were included in the latter. The lack of known higher order systems (on both smaller and larger scales) works against the MF and CSF, driving down the ratio.

\begin{figure*}[h]
    \centering
    \includegraphics[width=1.0\textwidth]{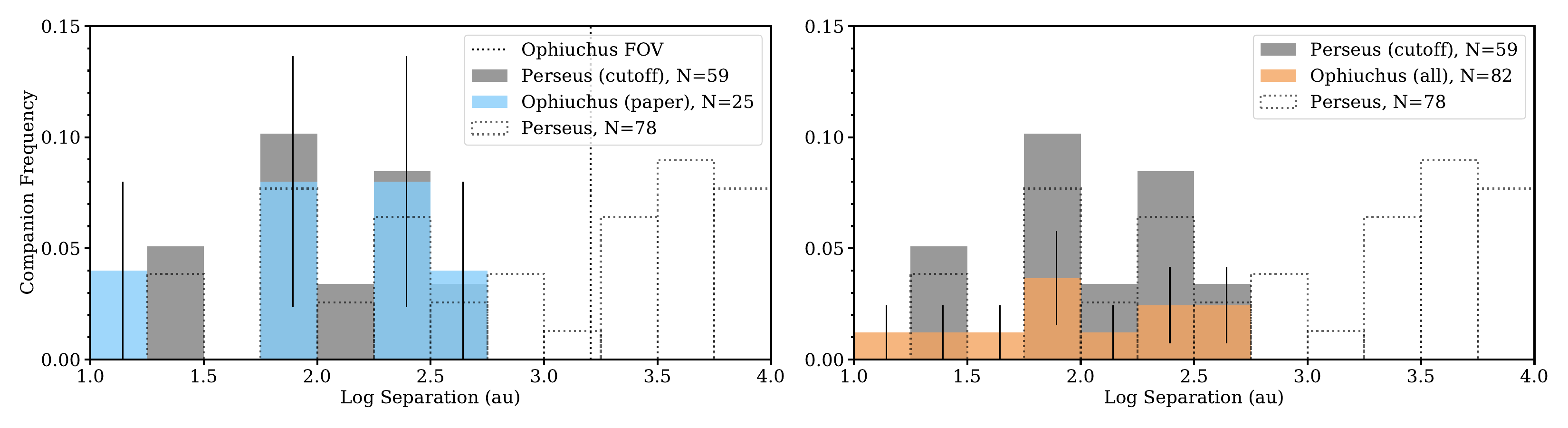}
    \caption{Histograms comparing the companion frequency between binaries in Perseus (Fig. 26, top left, full sample from \cite{tobin2018}) and Ophiuchus. {\bf Left:} The histogram shows the full Perseus binary sample outlined with dots and a subset of that sample, named "cutoff", in grey. Our Ophiuchus data, named "paper", are in blue with black error bars. The field of view of the Ophiuchus data are at 1600 au, represented by the vertical dotted line. {\bf Right:} Same as before but the Ophiuchus data contain addition binaries from \cite{Cieza2019}, VLA 1623 (\cite{looney2000}, \cite{hsieh2020}) and IRAS 16293 \citep{wootten1989} in orange.}
    \label{fig:binary_plot}
\end{figure*}

\begin{figure*}[h]
    \centering
    \includegraphics[width=1.0\textwidth]{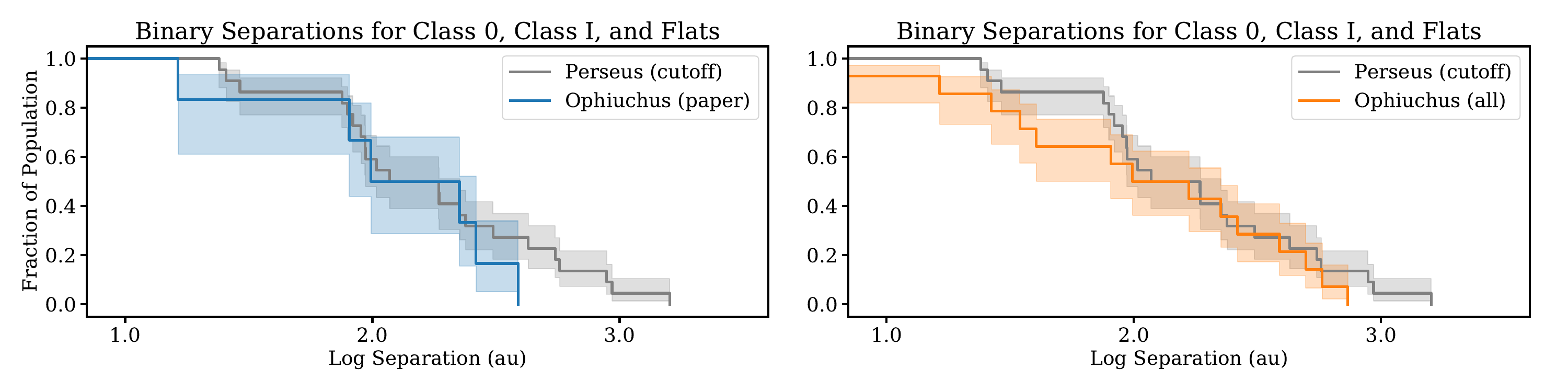}
    \caption{Cumulative distributions of binary separations in Perseus and Ophiuchus. {\bf Left:} We compare our sample (blue) to the "cutoff" sample from Perseus (grey). The Perseus sample has a maximum separation of 1000 au.  {\bf Right:} Same as before but the Ophiuchus data contain addition binaries from \cite{Cieza2019}, VLA 1623 (\cite{looney2000}, \cite{hsieh2020}) and IRAS 16293 \citep{wootten1989} in orange.}
    \label{fig:cdf__plot}
\end{figure*}

To delve further into why this was the case, we compared the multiplicity separations in Perseus and Ophiuchus (shown in Figure \ref{fig:binary_plot} and Figure \ref{fig:cdf__plot}). We compared the Perseus data with a cut-off at the Ophiuchus field of view limit. This was done primarily because it is ill-advised to compare the bimodal Perseus data with the unimodal Ophiuchus data. Furthermore, the bimodal nature of the Perseus data only reveal themselves at distances well outside what we could recover in a single pointing. It is also very likely that there are two different formation mechanisms reflected in the Perseus data.

Statistical analysis of these data consisted of an Anderson-Darling (AD) test. This mimics the analysis done by previous works in this field \citep[e.g.,][]{cox2017,tobin2018}. The AD test was chosen because it is more robust than the Kolmogorov-Smirnov test, namely in placing more weight at the end points of the distributions rather than the middle. We also avoided the log-rank test since our cumulative distribution functions crossed, a condition that causes the test to perform sub-optimally. We believe the conditions for using the AD test were fulfilled in our case. Its null hypothesis posits that the two distributions were drawn from the same distribution. We report that we failed to reject this hypothesis. When the "cutoff" Perseus data were compared with only the sources from this paper, our p-value was 0.753 and the AD statistic was 0.4975. When the cutoff Perseus data were compared with all the Ophiuchus data (our paper + \cite{Cieza2019} + IRAS 16293 + VLA 1623), we report a p-value of 0.779 with a AD statistic of 0.4577. Even though there does seem to be a difference in the young Ophiuchus disk masses compared to other regions, there does not seem to any difference in the multiplicity.

\section{Summary} \label{sec:summary}

We have conducted an 870 $\mu$m dust continuum survey of 31 protostars in the Ophiuchus region, focusing on the youngest members of the cloud (i.e., Class 0, I, and flat spectrum).  The spatial resolution of the observations was $\sim0.15^{\prime\prime}$ (21 au).  We detected 24 of the 25 sources in the pointing center and an additional 6 companions.  

Our main results are:

\begin{itemize}

\item Disk masses:  In our analysis, we used similar assumptions to those of \cite{tobin2020}.  The average disk masses of our Class I and Flat spectrum protostars are 2.8$^{+2.1}_{-1.2}$ and 2.5$^{+9.2}_{-1.1}$ M$_\Earth$, respectively. On average, our disk mass calculation assumptions give 50\% smaller disk mass estimates than the assumptions in \cite{williams2019}. This is largely due to the difference in temperature used for each calculation, which is dependent on source bolometric luminosity in our case. We combine our data with that of \cite{williams2019} to compare the masses of the Oph Class I and Flat spectrum protostars with those of Orion and Perseus.  As previously suggested by \cite{tobin2020}, the Oph sources have much lower mass disks even when using the same assumptions and continuum observing band. Although some of the difference may be due to misidentifying of Class, we posit that the overall trend, in conjunction with the variation in radii, arises from a fundamental difference in the initial conditions of the star formation process in the two clouds.

\item Disk sizes: We find mean radii of 23.5$^{+1.8}_{-1.2}$ and 16.5$^{+2.8}_{-0.9}$ au for Class I and Flat spectrum protostars, respectively, in our sample.  These mean radii were derived by fitting a Gaussian to the CDF from the Kaplan-Meier estimator.  When compared to the disk radii in the VANDAM Perseus sample and Orion samples, the Ophiuchus disks have smaller disk sizes, which is in agreement with the lower masses.  

\item Multiplicity: Although we see a difference in the disk masses and radii between Ophiuchus and Orion/Perseus, that is not necessarily the case with the multiplicity.
The MF and CSF multiplicity statistics are consistent between Perseus and Orion.  The degree of consistency increases when more complete samples are compared; however, further work is needed to better classify the C2D catalog source evolutionary classes. Finally, a comparison of the multiplicity populations of Perseus and Oph via an Anderson-Darling test fails to reject the null hypothesis that the two different star-forming regions were drawn from the same distribution. 

\item Although the multiplicity statistics of Ophiuchus is similar to other star forming regions, there are clear differences in the disk properties of Ophiuchus.   However, as we were mostly probing close binaries, which are thought to be formed via disk fragmentation \citep{tobin2016}, one would also expect to see a difference in the multiplicity.   Nonetheless, it seems possible that we are seeing some differences in the star formation process between Ophiuchus and Orion/Perseus that may be linked to the angular momentum transport processes, which would have implications with planet formation.  More study on the differences may be able to address these questions.

\end{itemize}

\section{Acknowledgements} \label{sec:ack}

This paper makes use of the following ALMA data: ADS/JAO.ALMA\#2015.1.00741.S. ALMA is a partnership of ESO (representing its member states), NSF (USA) and NINS (Japan), together with NRC (Canada), MOST and ASIAA (Taiwan), and KASI (Republic of Korea), in cooperation with the Republic of Chile. The Joint ALMA Observatory is operated by ESO, AUI/NRAO and NAOJ. The National Radio Astronomy Observatory is a facility of the National Science Foundation operated under cooperative agreement by Associated Universities, Inc.
LWL acknowledges support from NSF AST-1910364.
ZYL is supported in part by NASA 80NSSC20K0533 and NSF AST-1910106

\facility{ALMA}
\software{CASA \citep[https://casa.nrao.edu;][]{casa},
astropy (The Astropy Collaboration 2013, 2018)}

% Tab 1
\input{sourcetable_deluxe.tex}

% Tab 2
\begin{table*}[ph!]
    %\small
    \centering
    \caption{Binary Table} \label{tb:binaries}
    \input{septable.tex}
\end{table*}

% Tab 3
\begin{table*}[ph!]
    %\small
    \centering
    \caption{Multiplicity Fraction and Companion Star Fraction} \label{tb:mfncsf}
    \input{mfncsf_table.tex}
\end{table*}

% Fig 1

% Fig 2

% Fig 3

% Fig 4

% Fig 5

% Fig 6

% Fig 7

\end{document}

%% file: sourcetable_deluxe.tex
\setlength{\tabcolsep}{2.6pt}
\hskip-2.0cm

\onecolumngrid
\newpage
\centerwidetable
\movetabledown=2cm
%\movetableright=5cm
\begin{rotatetable}
\begin{deluxetable}{c c c c c c c c c c r r c c}

\tabletypesize{\footnotesize}
\tablecaption{Observed Sources \label{tb:sources}}

\tablehead{
    \colhead{Oph-} & \colhead{Class$^{a}$} & \colhead{RA} & \colhead{Dec} & \colhead{F$_{\nu}$$^{b}$} & \colhead{Peak F$_{\nu}$$^{c}$} &     \colhead{a$^{b}$} & \colhead{b$^{b}$} & \colhead{PA$^{b}$} & \colhead{Incl.} & \colhead{L$_{bol}$$^{d}$} & \colhead{M$_{dust}$$^{e}$} & \colhead{R$_{disk}$$^{f}$} & \colhead{Beam} \\[-0.2cm]
    \colhead{emb} & \colhead{} & \colhead{(J2000)} & \colhead{(J2000)} & \colhead{(mJy)} & \colhead{(mJy bm$^{-1}$)} & \colhead{(mas)} & \colhead{(mas)} & \colhead{($^{\circ}$)} & \colhead{($^{\circ}$)} & \colhead{(L$_{Sol}$)} & \colhead{(M$_{\Earth}$)} & \colhead{(au)} & \colhead{(mas)}
}
\startdata
 1  & 0     & 16:28:21.62   & -24:36:24.17  & 34.43$\pm$0.75    & 23.35$\pm$0.32    & 125.5$\pm$3.5 & 48.9$\pm$2.5  & 114.38$\pm$2.00   & 67        & 0.25      & 6.8       & 14.9$\pm$0.4  & 150x110   \\
 4  & I     & 16:31:36.78   & -24:04:20.34  & 29.43$\pm$0.88    & 22.50$\pm$0.38    & 255.9$\pm$17.1& 88.1$\pm$9.0  & 77.0$\pm$3.5      & 67        & 0.18      & 6.5       & 30.4$\pm$2.0  & 440x230   \\
 5  & I     & 16:27:21      & -24:27:27     & \nodata           & $\leq$0.66        & \nodata       & \nodata       & \nodata           & \nodata   & \nodata   & \nodata   & \nodata       & 150x110   \\
 6  & I     & 16:27:05.25   & -24:36:30.16  & 118.5$\pm$2.3     & 25.91$\pm$0.41    & 444$\pm$9     & 117$\pm$4     & 168.52$\pm$0.54   & 75        & 0.15      & 27.8      & 52.8$\pm$1.1  & 150x120   \\
 7  & I     & 16:31:52.05   & -24:57:26.36  & 1.50$\pm$0.29     & 1.24$\pm$0.15     & $\leq$84      & $\leq$6       & \nodata           & 86        & 0.10      & 0.4       & $\leq$ 10     & 150x110   \\
 8  & I     & 16:26:21.36   & -24:23:04.90  & 28.76$\pm$0.72    & 25.00$\pm$0.37    & $\leq$71      & $\leq$9.7     & \nodata           & 82        & 18.30     & 1.6       & $\leq$ 8      & 150x110   \\
 9  & I     & 16:26:25.47   & -24:23:01.83  & 99.31$\pm$0.98    & 39.10$\pm$0.29    & 238$\pm$3     & 101$\pm$2     & 26.98$\pm$0.63    & 65        & 0.10      & 26.8      & 28.3$\pm$0.4  & 150x130   \\
10 A& I     & 16:32:00.98   & -24:56:43.46  & 131.10$\pm$2.95   & 87.17$\pm$1.27    & 99.4$\pm$4.0  & 65.0$\pm$3.9  & 92.10$\pm$7.91    & 49        & 2.70      & 12.6      & 11.8$\pm$0.5  & 150x110   \\
10 B& I     & 16:32:00.99   & -24:56:42.76  & 18.12$\pm$1.29    & 16.76$\pm$0.70    & \nodata       & \nodata       & \nodata           & \nodata   & 2.70      & 1.7       & \nodata       & 150x110   \\
11 A& I     & 16:27:17.58   & -24:28:56.80  & 24.65$\pm$0.62    & 16.71$\pm$0.27    & 133.4$\pm$4.1 & 40.0$\pm$8.0  & 5.98$\pm$3.20     & 73        & 0.52      & 3.9       & 15.9$\pm$0.5  & 150x130   \\
11 B& I     & 16:27:17.44   & -24:28:56.55  & 2.55$\pm$0.47     & 2.21$\pm$0.24     & $\leq$87      & $\leq$54      & \nodata           & 52        & 0.52      & 0.4       & $\leq$ 10     & 150x130   \\
11 C& I     & 16:27:17.42   & -24:28:55.01  & 2.37$\pm$0.64     & 1.45$\pm$0.25     & $\leq$260     & $\leq$64      & \nodata           & 76        & 0.52      & 0.4       & $\leq$ 31     & 150x130   \\
12  & I     & 16:27:24.59   & -24:41:03.71  & 11.12$\pm$1.05    & 12.91$\pm$0.66    & \nodata       & \nodata       & \nodata           & \nodata   & 0.33      & 2.0       & \nodata       & 170x130   \\
13  & I     & 16:27:27.99   & -24:39:33.94  & 30.35$\pm$0.51    & 20.90$\pm$0.23    & 52.0$\pm$8.8  & 16.3$\pm$14.3 & 140$\pm$12        & 72        & 7.20      & 2.2       & 6.2$\pm$1.0   & 150x110   \\
14 A& I     & 16:27:26.91   & -24:40:50.70  & 34.47$\pm$1.37    & 25.14$\pm$0.63    & 90.7$\pm$6.5  & 40.2$\pm$6.0  & 126.16$\pm$7.54   & 64        & 3.80      & 3.0       & 10.8$\pm$0.8  & 150x110   \\
14 B& I     & 16:27:26.92   & -24:40:51.28  & 4.26$\pm$0.85     & 2.91$\pm$0.37     & \nodata       & \nodata       & \nodata           & \nodata   & 3.80      & 0.4       & \nodata       & 150x110   \\
15  & I     & 16:31:52.44   & -24:55:36.51  & 15.17$\pm$1.32    & 15.12$\pm$0.76    & \nodata       & \nodata       & \nodata           & \nodata   & 0.13      & 3.7       & \nodata       & 190x150   \\
17  & I     & 16:31:35.65   & -24:01:29.91  & 785.06$\pm$42.41  & 64.85$\pm$3.25    & 522$\pm$9.4   & 361$\pm$6.6   & 148.85$\pm$7.10   & 46        & 1.50      & 89.3      & 62.1$\pm$1.1  & 150x110   \\
18  & I     & 16:28:57.87   & -24:40:55.34  & 5.39$\pm$0.49     & 4.66$\pm$0.25     & 89$\pm$28     & 18$\pm$24     & 129.06$\pm$22.83  & 78        & 0.03      & 2.3       & 10.6$\pm$3.3  & 150x110   \\
19  & F     & 16:27:28.45   & -24:27:21.64  & 4.28$\pm$0.53     & 2.69$\pm$0.22     & 157$\pm$29    & 52$\pm$38     & 92.28$\pm$13.98   & 71        & 0.47      & 0.7       & 18.7$\pm$3.4  & 150x110   \\
20  & I     & 16:27:06.77   & -24:38:15.43  & 69.3$\pm$5.8      & 24.6$\pm$1.6      & 407$\pm$29    & 342$\pm$25    & 130.4$\pm$3.9     & 67        & 0.60      & 10.4      & 48.4$\pm$3.4  & 150x110   \\
21  & I     & 16:27:02.33   & -24:37:27.66  & 6.15$\pm$0.44     & 5.27$\pm$0.23     & $\leq$79      & $\leq$36      & \nodata           & 63        & 5.20      & 0.5       & $\leq$ 9      & 150x110   \\
22  & I     & 16:26:40.46   & -24:27:14.92  & 209.74$\pm$16.07  & 32.47$\pm$2.17    & 936$\pm$32    & 631$\pm$21    & 164.8$\pm$3.2     & 48        & 0.06      & 67.7      & 111.3$\pm$3.8 & 350x190   \\
23  & F     & 16:26:48.48   & -24:28:39.29  & 105.05$\pm$2.27   & 25.40$\pm$0.45    & 420$\pm$7.8   & 143$\pm$3.5   & 134.46$\pm$0.87   & 70        & 0.12      & 26.6      & 50.0$\pm$0.9  & 190x120   \\
24  & I     & 16:27:37.24   & -24:42:38.41  & 11.80$\pm$0.52    & 12.16$\pm$0.30    & \nodata       & \nodata       & \nodata           & \nodata   & 0.13      & 2.9       & \nodata       & 140x130   \\
25  & I     & 16:31:43.75   & -24:55:24.92  & 19.52$\pm$0.34    & 16.85$\pm$0.18    & 53.6$\pm$5.1  & 37.8$\pm$12.5 & 19$\pm$24         & 45        & 0.28      & 3.7       & 6.4$\pm$0.6   & 150x110   \\
26 A& F     & 16:27:30.18   & -24:27:43.80  & 12.86$\pm$0.49    & 12.90$\pm$0.28    & \nodata       & \nodata       & \nodata           & \nodata   & 0.93      & 1.7       & \nodata       & 160x150   \\
26 B$^{g}$& F?    & 16:27:30.17   & -24:27:43.88  & 9.22$\pm$0.53     & 8.51$\pm$0.29     & $\leq$59      & $\leq$28      & \nodata           & 62        & 0.93      & 1.2       & $\leq$ 7      & 160x150   \\
27 A& F     & 16:27:39.82   & -24:43:15.51  & 156.31$\pm$1.80   & 57.92$\pm$0.50    & 198.6$\pm$2.4 & 140.4$\pm$1.5 & 115.57$\pm$1.60   & 45        & 0.74      & 22.0      & 23.6$\pm$0.3  & 160x110   \\
27 B& F?    & 16:27:39.84   & -24:43:13.89  & 11.86$\pm$0.72    & 7.41$\pm$0.30     & 143$\pm$16    & 28$\pm$33     & 59.28$\pm$7.23    & 79        & 0.74      & 1.7       & 17.0$\pm$1.9  & 160x110   \\
28  & F     & 16:27:21.46   & -24:41:43.53  & 26.64$\pm$0.38    & 21.70$\pm$0.19    & 81.6$\pm$4.0  & 15.1$\pm$12.9 & 66.0$\pm$4.0      & 79        & 1.20      & 3.2       & 9.7$\pm$0.5   & 150x110   \\
\enddata
\tablenotetext{}{$^{a}$\cite{Cieza2019} where available, otherwise \cite{enoch2009}. $^{b}$Deconvolved Gaussian fit derived uncertainty using CASA's \texttt{imfit}. $^{c}$Measured with image uncertainty. $^{d}$Values obtained from \cite{enoch2009}. $^{e}$Calculated from Eq. \ref{eq:mdust}. $^{f}$Calculated using 2$\sigma$ of the Gaussian fit. $^{g}$Considered a new source in this paper.}
\end{deluxetable}
\end{rotatetable}

%% file: septable.tex
\begin{tabular}{l c r} 
\hline \hline
Binary Pair& Separation$^{a}$ & Ref.$^{b}$  \\ 
  & (au)  & \\
\hline
 Oph-emb 10 A - B           & 100.9     & 1\\
 Oph-emb 11 A - B           & 268.4     & 1\\
 Oph-emb 11 A - C           & 396.8     & 1\\
 Oph-emb 14 A - B           & 82.5      & 1\\
 Oph-emb 26 A - B           & 16.7      & 1\\
 Oph-emb 27 A - B           & 229.9     & 1\\
 J162636.8-241552           & 4.6       & 2\\
 J162648.5-242839           & 590.8     & 2\\
 J162715.4-242640           & 35.4      & 2\\
 J163152.1-245616           & 416.8     & 2\\
 VLA 1623 A - B             & 167.1     & 3,4\\
 VLA 1623 A1 - A2           & 26.5      & 4,5\\
 IRAS 16293-2242 A - B      & 734.2     & 6\\
 IRAS 16293-2242 A1 - A2    & 40.3      & 6\\
\hline
\end{tabular}
\tablenotetext{}{$^{a}$Higher order systems were reduced to binaries with source A as the common companion. $^{b}$All distances adjusted to 140 pc. $^{1}$This paper. $^{2}$\cite{Cieza2019}. $^{3}$\cite{looney2000}. $^{4}$\cite{harris2018}. $^{5}$\cite{hsieh2020}. $^{6}$\cite{wootten1989}}

%% file: mfncsf_table.tex
\begin{tabular}{l l c c r} 
\hline \hline
Sample$^{a}$ & S:B:T:Q & MF & CSF & Ref.  \\ 
\hline
 Ophiuchus Class I + Flats (Paper)          & 18:4:2:0      & 0.25 $\pm$ 0.09   & 0.33 $\pm$ 0.10   & 1\\
 Ophiuchus Class I + Flats (Paper + Cieza)  & 72:7:2:0      & 0.11 $\pm$ 0.04   & 0.14 $\pm$ 0.04   & 1,2\\
 Perseus Class I + Flats                    & 21:8:0:0      & 0.28 $\pm$ 0.08   & 0.28 $\pm$ 0.08   & 3\\ 
 Orion Class I + Flats                      & 182:32:2:0    & 0.16 $\pm$ 0.02   & 0.17 $\pm$ 0.03   & 3\\
 
\hline
\end{tabular}
\tablenotetext{}{$^{a}$The sampled maximum separation range is 1000 au. $^{1}$This paper. $^{2}$\cite{Cieza2019}. $^{3}$Tobin et al. (2021, in preparation) in the 20-1000 au range using their probabilistic weighting which corrects for more crowded regions.}